\documentclass{astlb}
\usepackage{longtable}
\usepackage{lscape}
\usepackage{graphicx}
\usepackage{color}
\usepackage{setspace}
\usepackage{amsmath}

\usepackage{hyperref}

\definecolor{darkblue}{rgb}{0,0,0.9}

\def\ps1{\emph{Pan-STARRS1}}

\begin{document}

\journalinfo{2023}{0}{0}{1}[0]

\title{Transient events in the near-nuclear regions of AGNs and quasars as the sources of the proper motion imitations }

\author{I.~M.~Khamitov\email{irek\_khamitov@hotmail.com}\address{1,2},
  I.~F.~Bikmaev\address{1,2},
   M.~R.~Gilfanov\address{3,4},
  R.~A.~Sunyaev\address{3,4},
  P.~S.~Medvedev\address{3},
  M.~A.~Gorbachev\address{1,2}\\
$^1$\it{Kazan Federal University, Kazan, Russia\\}
$^2$\it{Tatarstan Academy of Sciences, Kazan, Russia\\}
$^3$\it{Space Research Institute of Russian Academy of Sciences, Moscow, Russia\\}
$^4$\it{Max-Planck Institute for Astrophysics, Garching, Germany}
}

\shortauthor{I.~M.~Khamitov et al.}

\shorttitle{Transient events in the near-nuclear region of AGNs}
 
\submitted{\today}


\begin{abstract}
This paper is an extension of the \cite{2022AstL...48..724K} study in terms of cataloging and astrophysical interpretation of imitation of significant proper motions in galaxies with active nuclei and quasars which have been measured using data from the GAIA space observatory. 
We present a sample of SRG/eROSITA X-ray sources located in the eastern Galactic hemisphere ($0^\circ<l<180^\circ$), 
with significant proper motions according to GAIA eDR3 measurements and whose extragalactic nature has been confirmed.   
The catalog consists of 248 extragalactic sources with spectroscopically measured redshifts. It includes all objects available in the Simbad database and matched to the identified optical component within a radius of 0.5 arcsec.  Additionally, the catalog includes 18 sources with the spectral redshift measurements based on observations at the Russian-Turkish 1.5-m telescope RTT-150, \cite{2022AstL...48..724K}. 
The sources of the catalog are AGNs of various types (Sy1, Sy2, LINER), quasars, radio galaxies, and star-forming galaxies. 
The imitation of significant proper motions can be explained (previously known in astrometry as the VIM effect) by the presence of transient events    
on the line of sight in the field of view of AGN nuclei and quasars (within the GAIA resolution element). Such astrophysical phenomena may be the supernovae outbursts, tidal destruction events in AGNs with double nuclei, variability of large-mass supergiants, the presence of O-B associations in field of view of variable brightness AGN, etc. 
A model of flares with a fast rise  and exponential decay profile allows to describe the variable positional parameters of most similar sources observed in GAIA.  This cross-matching approach of  the X-ray source catalogs of the SRG/eROSITA observatory and the optical catalog of the GAIA observatory can be used as an independent technique for detecting transient events in the neighborhood of AGN core (on scales of several hundred parsecs in the picture plane).

\end{abstract}

\section{Introduction}

The \cite{Klioner2022}, \cite{LCAQ5}, \cite{2022ApJ...933...28M}, and \cite{2022AstL...48..724K} presented in their papers lists of quasars - point-like objects, 
not resolved by the GAIA satellite but are extragalactic according to their observed redshifts, with large pecular motions according to the GAIA astrometric measurements \citep{GAIAEDR3}. In the extragalactic sources reference frame the observed amplitudes of proper motions correspond to displacements of matter with velocities exceeding the speed of light by tens and hundreds of times. The imitation of the observed proper motions of AGN and quasars is most likely a manifestation of the well-known VIM effect, variability-induced-movers, caused by source variability. Initially, a method for investigating the implications of this effect was developed to analyze data of the space astrometry mission Hipparcos \citep{1996AA...314..679W}. It was focused to search for double stars in unresolved images, assuming that either one of the sources has variable brightness or there is a noticeable orbital motion in the binary system over the lifetime of the mission. The unprecedented positional accuracy, all-sky depth of the survey, and long duration of stable performance of the GAIA mission led to detection of numerous VIM occurrences in extragalactic sources in the optics. The observed amplitudes of proper motions of AGN and quasars are several orders of magnitude larger than the cases when the changes in the photocenter occur 
due to large-scale changes in the accretion disk and dust torus surrounding the central black hole, 
or due to the influence of primary gravitational waves or anisotropic expansion of the Universe, as well as microlensing (see \cite{LCAQ5} and references therein). In  \cite{2012AA...538A.107P} it is considered a model of a relativistic disk around a supermassive black hole (SMBH) including perturbations, 
leading to the brightening of a part of the disk and, consequently, to the displacement of the photocenter position. It was shown that the result of the rearrangement of the internal structure of the accretion disk can produce a photocenter offset of up to several mas (milli arcsec)  in the GAIA data, but only for bright quasars at low redshifts. Also in this paper, the proposed model was applied to long-term observations of a sample of 20 quasars exhibiting significant photocenter variability. For the cases of SDSS J121855.80+020002.1 and Mrk 877, the model was insufficient and the possibility of supernovae exploding very close to the central AGN source and a possible indication of a kpc (pc) scale binary SMBH was discussed.  In \cite{2022ApJ...933...28M} a comparison of the MIRAGN and GAIA eDR3 catalogs resulted in a list of 44 candidate double and multi-system quasars with proper motions and 4 known gravitationally-lensed systems. It is suggested that many proper motion quasars may be closer unresolved double systems exhibiting the VIM effect, and a smaller fraction may be random coincidences with foreground stars causing weak gravitational lensing. The \cite{2023MNRAS.522.1736P} shows that galaxy interactions are the dominant mechanism for triggering quasar activity in the local Universe. It is shown that the host galaxies of type II quasars in $\sim65\%$ show morphological features corresponding to mergers or encounters of galaxies. In contrast to the idea that quasars are triggered at galaxy merger peaks, when two nuclei coalesce, and only become visible post-coalescence, 
most of the morphologically perturbed type II quasars in the \cite{2023MNRAS.522.1736P} sample are observed in the pre-coalescence phase ($61^{+8}_{-9}\%$). 

The detection of AGN with source position offset on scales up to several $mas$, measured from VLBI radio data relative to GAIA measurements in the directions along and opposite to the jets, indicates the presence of strong extended optical jet structures in these systems on parsec scales \citep{2017AA...598L...1K}.  These studies have shown that  to explain the VLBI-GAIA bias the radiative sizes of the required strong optical jet  should be at least 20-50 pc \citep{2019ApJ...871..143P}.  Differential brightness variations in jets of this type can lead to the observed VIM effect in GAIA data. Optically bright jets with ultrarelativistic proper motions were also observed in the nearest strong radio galaxy M87 \citep{1999ApJ...520..621B}, 
where the Hubble telescope was able to observe motions of details in the image of jets with apparent velocities up to 6-8 light speeds. \cite{LCAQ5} briefly investigated both of the above scenarios and concluded that radio jet activity cannot be the main factor affecting the apparent proper motion in quasars in optics.

In \cite{2022AstL...48..724K} we report a sample of 502 extragalactic X-ray sources detected by the eROSITA telescope \citep{2021AA...647A...1P}  of the space observatory Spectrum-Roentgen-Gamma \citep{2021AA...656A.132S} in the eastern galactic hemisphere, for which the Russian consortium of eROSITA telescopes is responsible for data processing ($0^\circ<l<180^\circ$), and for which the GAIA satellite has measured significant values of proper motions. Most of these sources belong to extended optical sources according to GAIA, and only about 1.5\% of them are stars in our Galaxy. Using the Simbad database, it has been shown {\citep{2022AstL...48..724K} that an extragalactic nature is confirmed for about half of these 502 sources. For the remaining half, additional spectroscopic observations are required to estimate their redshifts and to do their optical identification. This work is carried out by the authors of the paper - based on observations on the 1.5-meter optical Russian-Turkish telescope RTT-150 (T\"UB\.ITAK National Observatory, Antalya, Turkey), the extragalactic nature of 18 sources has already been previously confirmed from the second half of the \cite{2022AstL...48..724K} list. In the current paper we present a catalog of 248 X-ray sources of eROSITA discussed in \cite{2022AstL...48..724K} 
for which spectroscopic redshift determinations and optical classification are available. The catalog includes all objects available in the Simbad database that match the identified optical component within a radius of 0.5 arcsec, and from spectral observations performed with the RTT-150 telescope. An additional source selection criterion was their absence among GAIA DR3 objects with spurious time-series signal by position related to the time-dependent angle of scanning of objects by GAIA detectors \citep{2023AA...674A..25H}. The catalog can be used in the future to search for and analyze the origin of GAIA significant proper motions of extragalactic objects from which formally follow the motions of matter in these sources with significant excesses of the speed of light.  

In the following tables we provide an additional list of 12 quasars, 2 blazars, and 4 radio galaxies from the X-ray catalog of the SRG/eROSITA telescope, which have large pecular motions according to the GAIA satellite data. Many blazars and part of quasars demonstrate the presence of strong optical jets. 

However, our catalog of extragalactic objects with large proper motions includes not only quasars or strong AGNs and radio galaxies, but also other systems. For example, galaxies with strong star formation, the nature of which is hardly related to the presence of powerful optical jets, are of independent interest. It may well turn out that in these objects with frequently exploding supernovae, the measured large apparent velocities of the central region of the galaxies are explained by a displacement of the galaxy's photocenter due to a simultaneously recorded bright supernova outburst at an appreciable distance from the galaxy's brightness center.  It is possible that for the same reason (bright supernova outburst in the surrounding galaxy) the brightness center of the galaxy with AGN can be shifted for a while, which is presented in the results of GAIA observations. 

The phenomenon of tidal destruction of stars by a supermassive black hole can also make the accretion disk around the black hole brighter than the whole galaxy for a distant observer for a time of the order of a year. Due to many reasons it will result in a shift of the measured photocenter. Moreover, for double systems it will lead to essential values. It is obvious that at such strong energy generation the bright optical jets with ultrarelativistic velocities can sometimes be formed. 

Below we have categorized the objects in our catalog by type so that professionals can more quickly find the objects they are interested in. The imitation of strong proper motion of extended extragalactic objects detected by GAIA may lead to interesting and unexpected conclusions about the physical processes resulting in such imitation.

To describe the observational parameters of the GAIA catalog - astrometric noise (astrometric\_excess\_noise) and the ratio of the total source displacement to this noise, we considered the effect of a FRED profiled flare on the position of the photocenter which took place at a given distance from a galaxy core.

\section{Catalog}

Spectroscopic redshift estimations are available for 248 objects. The catalog is presented as one common identification Table 1 and 9 Tables 2 - 10 with X-ray characteristics, classified by common object types: quasars - 11 sources; blazars - 2 sources; radio galaxies (Galaxy R) - 4 sources; Seyfert type 1 galaxies - 106 sources; Seyfert type 2 galaxies - 32 sources; LINER type AGNs - 11 sources; AGNs of undetermined type - 25 sources; star forming  galaxies (SF) - 15 sources and undetermined type galaxies - 42 sources. 
The identification table is ordered by right ascension, while within the X-ray tables, objects are listed in descending order of proper motion modulus according to GAIA eDR3 data.

The identification table has 9 columns:
1) SRG/eROSITA catalog identification number; 2) GAIA eDR3 catalog identification number; 3) RA (J2000.0); 4) DEC (J2000.0); 5) proper motion module with its measurement error in $mas$ units; 6) redshift; 7) source type; 8) identification in the Simbad database; 9) reference code for the article from which the redshift information was received. 

The X-ray characterization tables have 9 columns: 1) RA (J2000.0); 2) DEC (J2000.0); 3) magnitude in the G band (Gaia eDR3); 4) proper motion modulus in $mas$ units; 5) logarithm of the ratio of the X-ray flux $F_X$ from eROSITA data in the range 0.3--2.3 keV to the optical flux $F_{opt}$ in the G band from Gaia eDR3 data; 6) $X_{v}$ X-ray variability, defined as the ratio between the maximum and minimum flux values in the 4 eROSITA surveys without taking into account the flux measurement error; 7) X-ray luminosity from eROSITA data in the range 0.3--2.3 keV in its own reference frame without correction for internal absorption and absorption in the Galaxy\footnote{ For the calculation we assumed the standard cosmological model $\Lambda$CDM with the following parameters: $\Omega_m = 0. 3$, $\Omega_{\Lambda}$=0.7, $H_0 = 70$ km/s/Mpc}; 8) redshift; 9) lower estimation of the largest optical luminosity of the transient event falling within the GAIA catalog interval (see Appendix 1).

\section{Transient events as sources of proper motion imitation in AGN.}

Within the GAIA resolution element (60 $mas$) a transient event comparable in flux to the optical radiation of the AGN itself can be captured together with the AGN image.  In this case, due to a noticeable brightness variations of the transient event, the AGN photocenter registered by GAIA will vary in time and, consequently, a spurious signal of the AGN proper motion will be observed. Thus, in the projection onto the picture plane at a radius of 60 $mas$ from the apparent center of the AGN, there may be an additional sufficiently bright source located inside the host galaxy even at significant distances from the active nucleus. Such a source is most likely in the foreground because of the strong absorption near the core.

The distribution of the catalog's AGNs depending on the object type in the plane of astrometric noise ($\epsilon$, astrometric\_excess\_noise) and the ratio of the total source displacement to the astrometric noise ($2.8\mu/\epsilon$) according to the GAIA data is shown in Fig.\ref{fig:mu_epsi}. The astrometric noise parameter in GAIA is determined by the discrepancy between the source coordinates in the 5-parameter model of the global measurements and the photocenter measured directly on the base of individual observations as the weighted center. For nearby galaxies ($z<0.3$), the discrepancies are affected by the asymmetric structure of the galactic component due to the tracing of the GAIA detector windows at different angles. It is important that our catalog sources show no spurious signal of the temporal position dependence associated with the scan angle \citep{2023AA...674A..25H}. In addition, investigations of the offsets in the positions of VLBI to the GAIA photocenters in AGN have shown that in 73\% of the cases of significant offsets there were coincidences with the direction of the radiojet. This indicates that the host galaxy does not play an important role in the detected offsets and, therefore, argues in favor of a symmetric optical brightness distribution in the near-nuclear region  \citep{2019ApJ...871..143P}. Obviously, the discrepancies associated with a transient event will take place in the direction of the photocenter of the source in the absence of a transient to the position of the latter. Thus, for cases of close and bright AGN of the catalog, the level of astrometric noise not related to the VIM effect due to the transient event can be estimated from GAIA data by the cross-directional deviations of the photocenter with respect to the proper motion vector. The dotted lines are  the modeling results ($\mu,\epsilon$) for the case when the transient brightness variations corresponds to a fast rise  and exponential decay profile (FRED) that occurred at different apparent distances of the transient from the nucleus ($X_T$): black line - 60 $mas$, red line - 30 $mas$ (see Appendix 1). The distributions are constructed for the case of the ratio of the peak transient flux to the flux from the nucleus $R_0=1$, the fast rise parameter $\sigma=10^d$, and the decay factor $\alpha=111^d$, which corresponds to the decay $Co^{+56}$ characteristic of supernova explosions.

 \begin{figure*}
  \centering
  \includegraphics[width=0.66\columnwidth,height=0.55\columnwidth]{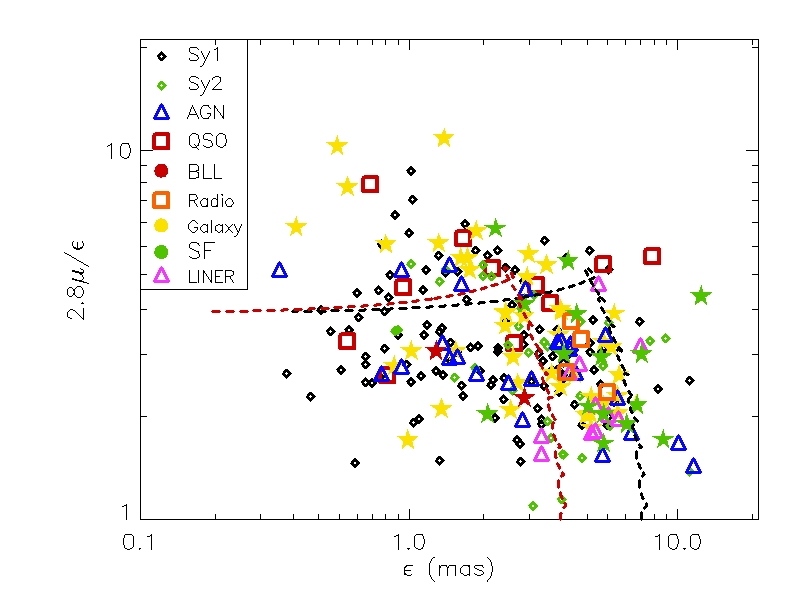}
  \includegraphics[width=0.66\columnwidth,height=0.55\columnwidth]{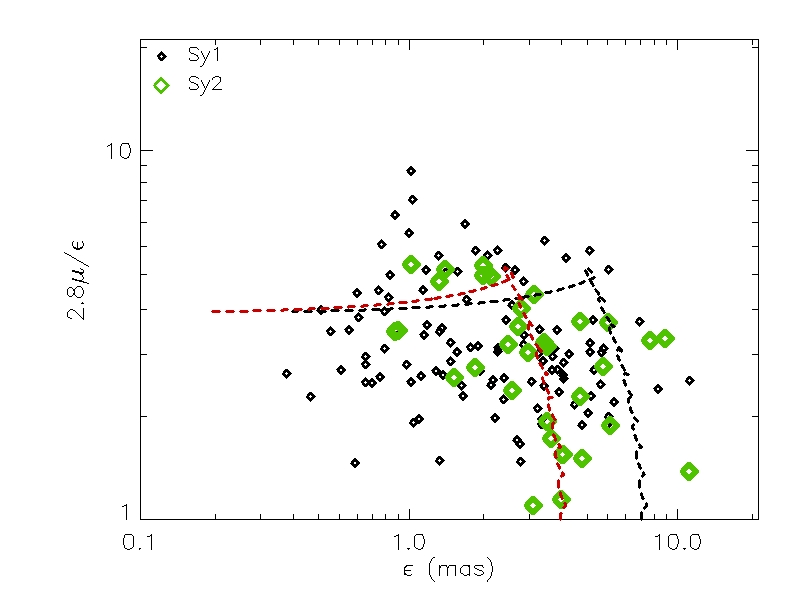}
  \includegraphics[width=0.66\columnwidth,height=0.55\columnwidth]{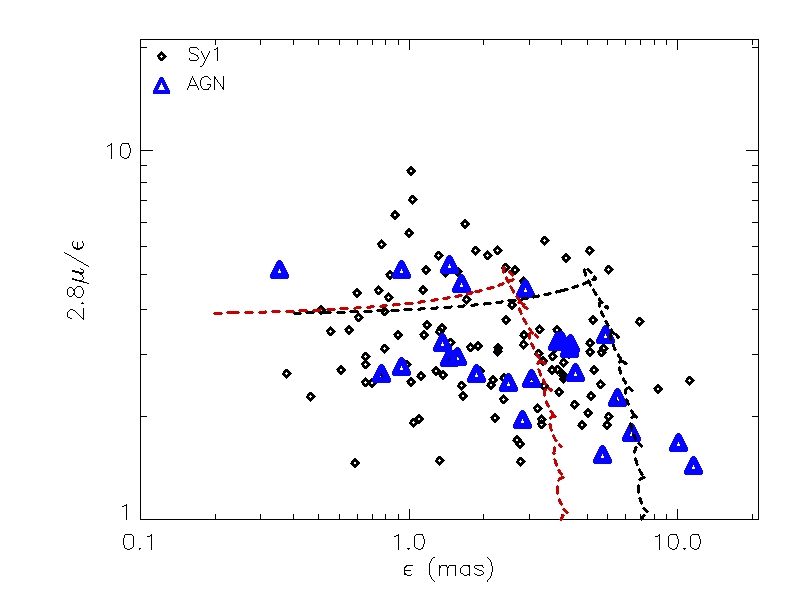}
  
  \includegraphics[width=0.66\columnwidth,height=0.55\columnwidth]{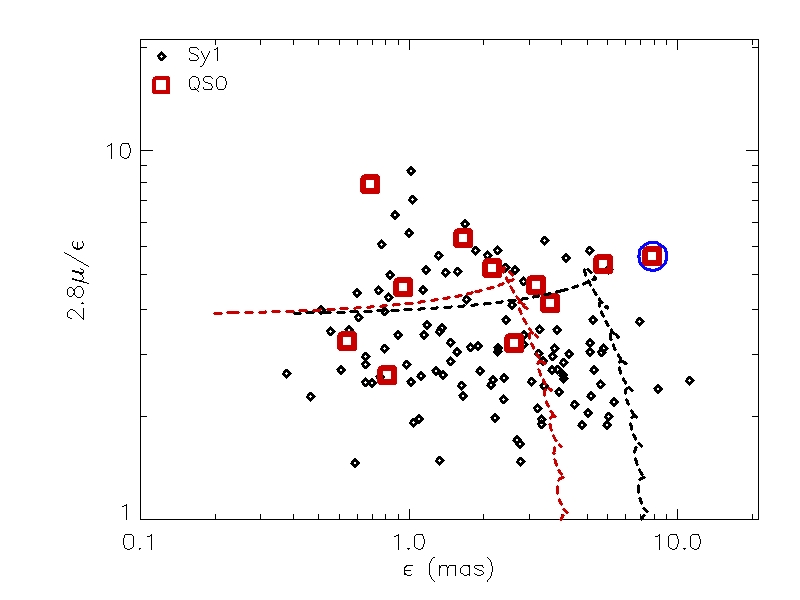}
  \includegraphics[width=0.66\columnwidth,height=0.55\columnwidth]{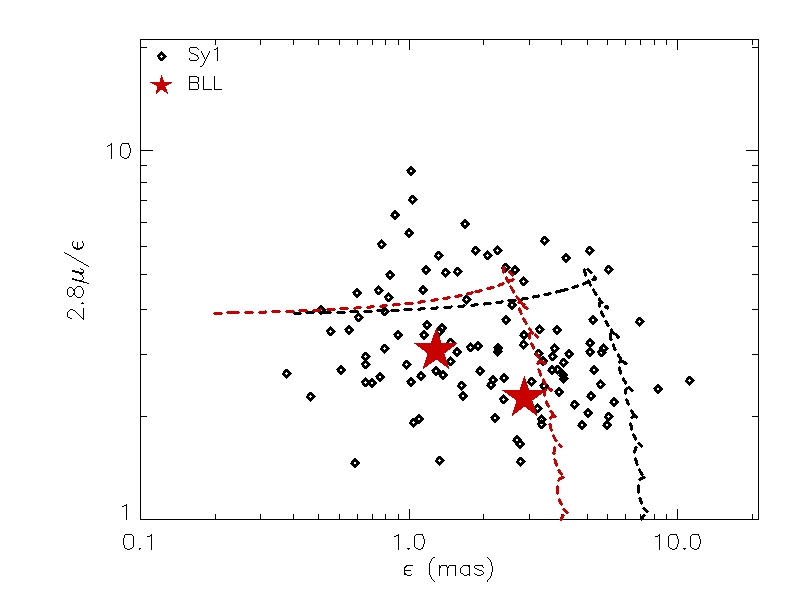}
  \includegraphics[width=0.66\columnwidth,height=0.55\columnwidth]{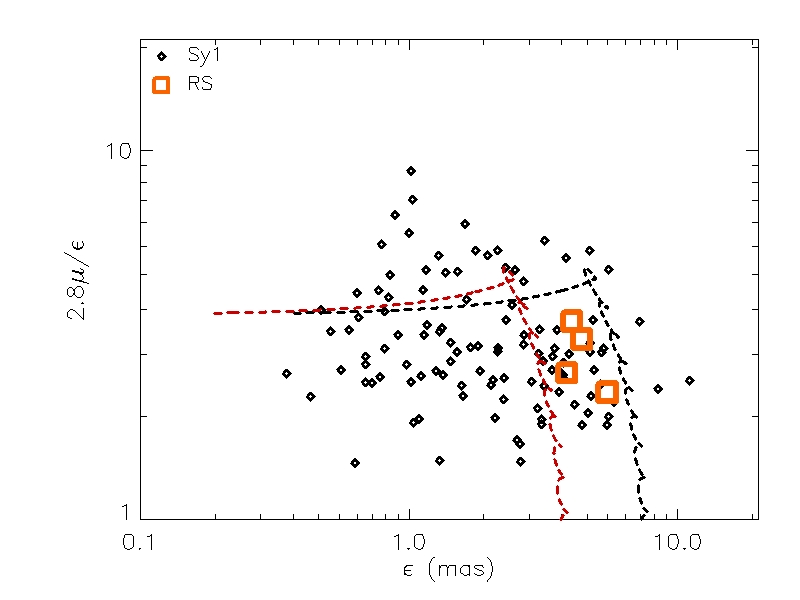}
  
  \includegraphics[width=0.66\columnwidth,height=0.55\columnwidth]{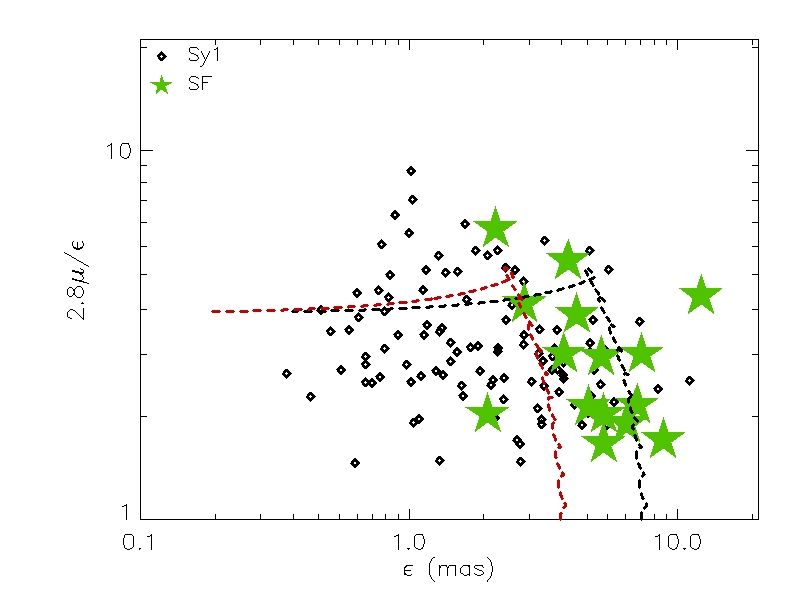} 
  \includegraphics[width=0.66\columnwidth,height=0.55\columnwidth]{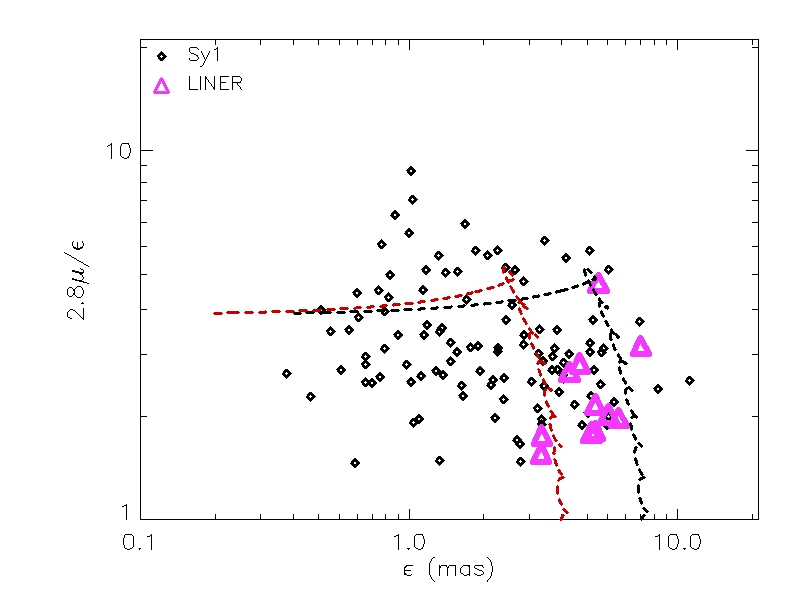}
  \includegraphics[width=0.66\columnwidth,height=0.55\columnwidth]{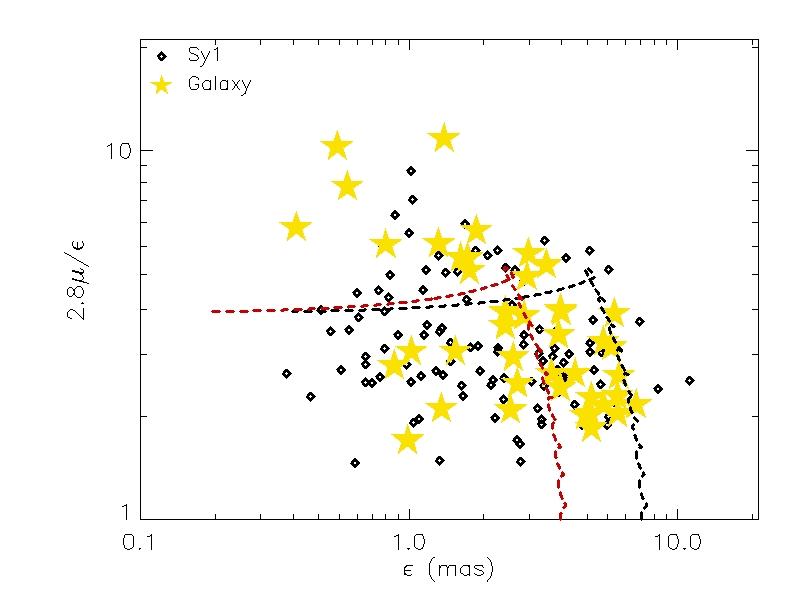}
    
  \hspace{1pt}
  \caption{Source type distribution in the plane $\epsilon - 2.8\mu/\epsilon  $.}
  \label{fig:mu_epsi}
  
\end{figure*}

The considered fitting of the observed parameters ($\mu,\epsilon$) by FRED flares is a multi-parameter problem that also depends on the cadence of source observations in GAIA.  Using optical light curves of GAIA sources from the AGN catalog with significant proper motions will provide model parameters.  Besides, the proposed model does not consider the spectral energy distribution (SED) of flares and its evolution with time. The necessity of SED including into the analysis becomes relevant depending on the redshift, since as the distance to the object increases, the recorded with the GAIA detectors bandwicth will correspond to more and more short-wavelength regions of the flare electromagnetic emission. To detailed modeling with SED it is required the complete source information of GAIA eDR3 catalog (photometric light curves and positional data) for the entire observation period. But a series of qualitative conclusions can already be drawn:
\begin{enumerate}
    \item 
    {
    On the GAIA light curves, in those cases where the maximum of the supernova outburst took place during the catalog interval, a relatively sharp and short-term brightness variation - an outlier - should be observed. The transient position with respect to the nucleus corresponds to the direction of the proper motion vector. But in cases where the observations were performed on the fading branch of the flare, the slow decrease in brightness can be interpreted as an intrinsic variability of the nucleus. The position of the transient is opposite to the direction of the proper motion vector with a slow decrease in modulus in contrast to the impact of the internal variability of the nucleus structure, which GAIA does not detect. Thus, it is possible to limit the time domains of the location of the flare maximum moment - before or during the GAIA catalog, i.e., horizontal or vertical solutions of the model.  
    }

    \item 
{
Using new astrometric data the reversal of the sign of the AGN apparent proper motion can be measured  for the case of the flare maximum located within the GAIA eDR3 catalog. The change in sign will take place in the subsequent GAIA observing period, when only the fading branch of the flare is observed.
}

 \item 
{
In galaxies with intense star formation, LINER, and radio galaxies, the transient event  is observed at distances of at least 30 $mas$ in the model with $R_0=1$. For models with $R_0<1.0$, or $\alpha < 111^d$ this lower limit is even higher. This limitation is on the one hand explained by the strong dust absorption in the near-core region in star-forming and LINER galaxies, but on the other hand it is an indirect indication that regions with a high-energy transient event should be closely related to the core activity.       
}

    \item 
{
The \cite{2023MNRAS.522.1736P} shows that galaxy interactions are the dominant mechanism for triggering quasar activity in the local Universe. It is shown that host galaxies with type II quasars in $\sim66\%$ demonstrate morphological features corresponding to mergers or encounters of galaxies. In contrast to the idea that quasars are triggered at galaxy merger peaks, when two nuclei merge, and become visible only after complete merging, most morphologically perturbed type II quasars in the sample from the \cite{2023MNRAS.522.1736P} paper are observed at a stage prior to the complete merging phase ($61^{+8}_{-9}\%$). Thus, the quasars in our sample with significant proper motions are very likely to be double nuclei within 60 $mas$ or on the order of a few hundred parsecs away. For quasar SRGeJ131118.5+463502, the lower estimation of the luminosity of the transient event within the GAIA catalog interval ($L_T$) is about $1.3\times10^{44} erg/s$ (in Fig. \ref{fig:mu_epsi} in the panel with quasars, this source is highlighted by a blue circle). This luminosity exceeds the maximum luminosity values for supernova events and is a candidate for a tidal disruption event (TDE) in a double-core system.  At the picture plane, the two SMBHs are separated by an angular distance of at least 60 $mas$, which at redshift z=0.271 corresponds to distances of the order of 250 parsecs.  Of course, to model the TDE it is necessary to use a more sophisticated dependence than FRED, involving an exponential function at large times after the flare maximum\citep{2021ApJ...908....4V}. The strongly variable AGNs found in the eROSITA catalog \citep{2022AstL...48..735M} may exhibit similar phenomena in the registration of their proper motion in the GAIA catalog. The VIM effect will manifest itself in the period when this variability has been detected, i.e., using GAIA data only in the period 2020-2022. 
}

\item 
{
To explain the apparent proper motion via a transient event in the neighborhood of the AGN nucleus, there are cases with apparent proper motions of gas ejections similar to those observed in the nearest strong radio galaxy M87 \citep{1999ApJ...520..621B}. The Hubble telescope made it possible to observe moving details in the image of a jet with apparent proper motions ($\mu_{jet}$) on the order of 20 $mas/yr$ at a relative brightness of $R\sim 0.01$ with respect to the optical core of M87. The model considered in Appendix 1 with linear transient motion based on the GAIA-measured proper motions of blazars from the  Table \ref{cat_BLL}  gives estimates of the jet brightness, at 20 $mas/yr$ proper motions of the jet, on the order of 10\% of the core brightness ($R=0.1$).  This value is an order of magnitude higher in relative luminosity compared to the optical details of the jets in M87, and the apparent magnitudes of the objects themselves are at the limit of detection by the GAIA detectors. But, on the other hand, in the experience of the M87, the GAIA resolution element may capture a large group of more faint details of the jet.   Together, they can provide a higher ratio ($R=0.1$). 

The presence of an optical jet in these objects may well explain the apparent proper motions. The presence of a flare in the neighborhood of the nucleus is not excluded in the cases of blazars as well. The use of astrometric trajectory from GAIA data allows us to separate the case of relativistic jet motion from a strong transit event. In the second case, variations in the position of the source will resemble the light curve of the flare (Fig.\ref{fig:lagm500},\ref{fig:lagp100}), as observed, for example, in the Kepler Observatory data in the search for double unresolved systems \citep{2016ApJS..224...19M}.
}

\item 
{
AGNs of undetermined type have a clear binary distribution at $2.8\mu/\epsilon$. This behavior is also evident in the Sy1 and Sy2 distributions. Combining sources of this type gives maximum values: $2.6\pm0.7$ and $4.7\pm0.5$. It should be noted that the larger value is approximated by the exponential decay parameter characteristic of supernova events. 
}

\item 
{
In the case of galaxies where the luminosity of the nucleus is not so high, in some cases the apparent proper motions can be modulated by long-period variations in the luminosity of stellar objects. For example, in the well-known double system of our Galaxy $\eta$ Carinae, quasi-periodic luminosity variations of several orders of magnitude over 5 years, reaching luminosity up to $10^{39}$ erg/s, have been observed \citep{1996ApJ...460L..49D}. Such cases can be determined by examining the GAIA catalog on different time samples, separated by several non-overlapping years. They will demonstrate long-period proper motion variability.
}

\item 
{
For tidal disruption events in the gravitational field of a supermassive black hole, initially the photocenter must not coincide with the nucleus. These can be double core cases. At core luminosities below $10^{40}$ erg/s, the initial offset of the photocenter can be provided by stellar O-B associations whose integral optical luminosity can be comparable to the core luminosity.
}

\item 
{
If our astrophysical interpretation of the proper motion imitation in the GAIA data is correct, then after the brightness of the transient event drops below the GAIA photometric threshold the extragalactic sources  will be observed at the permanent positions in subsequent GAIA measurements. Or the directions and amplitudes of the spurious proper motions will change, in case new transient events occurred in these AGNs during the next GAIA scans in subsequent mission periods, after 2017.
}

\end{enumerate}

\
 \begin{figure}[t]
  \centering
  \includegraphics[width=1\columnwidth]{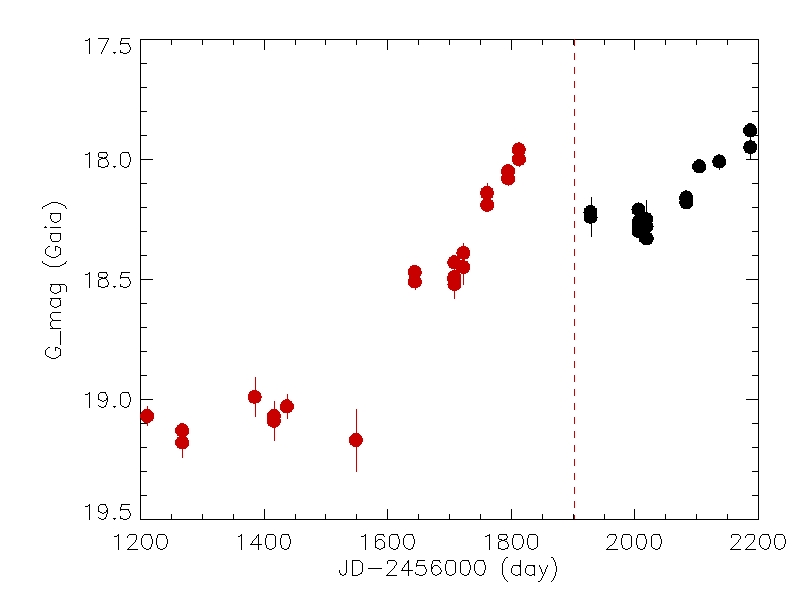}
    
  \hspace{1pt}
  \caption{Optical light curve of SRGeJ143701.5+264019 (quasar). Red circles show data corresponding to the period of the GAIA eDR3 catalog.}
  \label{fig:QSO}
  
\end{figure}

\section{Comparison with the GNT candidates catalog}

\cite{2018MNRAS.481..307K} presented a catalog of about 480 candidates for transient events in the near-nuclear regions of galaxies detected from GAIA photometric data and the SDSS DR12 galaxy catalog (GNT catalog). Among the spectrally confirmed sources, two matches with GNT were found: SRGeJ143701.5+264019 and SRGeJ151721.7+465812. The light curve of the source SRGeJ143701.5+264019 known as a quasar is shown in Fig.\ref{fig:QSO}. The data corresponding to the GAIA eDR3 catalog period is indicated with red circles (vertical dashed line is the upper boundary of the catalog).
\begin{figure}
  \centering
  \includegraphics[width=1\columnwidth]{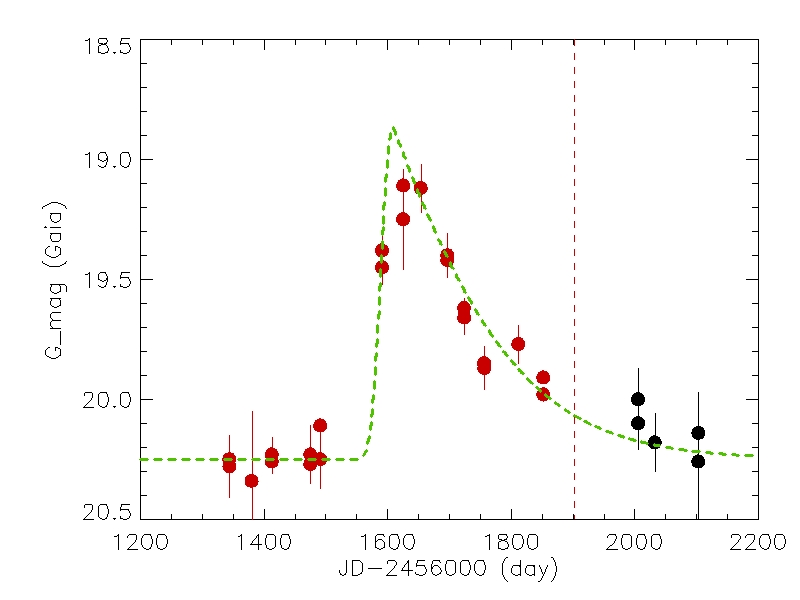}
    
  \hspace{1pt}
  \caption{Optical light curve of SRGeJ151721.7+465812 (star-forming galaxy). Red circles represent data corresponding to the period of the GAIA eDR3 catalog. Green dashed line shows the fit of the flare with FRED profile.}
  \label{fig:SN}
  
\end{figure}
 A monotonic rise in brightness is observed which continued even after the end of the GAIA eDR3 catalog with an approximately twofold drop in flux between 1800 and 2000 days on the graph. To obtain the flare profile the detailed positional information is required. The SRGeJ151721.7+465812 light curve has a completely different appearance where a flare with a FRED profile definitely stands out. The light curve is shown in Fig.\ref{fig:SN} and the green dashed line represents the fit of the flare with model parameters: $t_0=745^d,~R_0=2.6,~ \sigma=15^d,~ \alpha=111^d$. The proper motion ($17.63 \pm 2.3~ mas/yr$) and astrometric noise ($12.2~mas$) measured in GAIA, together with the flare parameters, are consistent with a transient distance from the core of $X_T \sim 80~mas$. Using the source redshift $z=0.154$ determined from the RTT-150 spectral data, we obtain that in the object reference frame the transient event took place at a distance of $\sim$214 pc. An estimation of the luminosity of the flare at maximum corresponds to $L_{max}=2.2 \times 10^{43}$ erg/s, which together with the exponential decay indicates that a Supernova event has occurred.

\section{Conclusion}
  
 The IMITATION of significant and high proper motions of AGN and quasars (identified by the SRG/eROSITA telescope as sources of X-ray emission) detected by GAIA measurements may be a manifestation of classical transient astrophysical events. These events take place in the near-nuclei regions (in the picture plane within the GAIA optical resolution element) of this type of extragalactic sources during long (several years) periods of high-precision measurements of their optical coordinates. Transient events resulting in displacements of AGN and quasar photocenters should have optical luminosities exceeding or comparable to the optical luminosities of AGN and quasars themselves. In our opinion, the most characteristic of the known astrophysical phenomena of such transient events are: supernovae outbursts, tidal disruption events in double SMBHs, high-amplitude long-term variations (increases or decreases) in the brightness of massive supergiant stars with outflowing envelopes and stellar winds of the $\eta$ Carinae type, as well as the presence of stellar O-B associations in the field of view of AGN of variable brightness.  This interpretation is more astrophysically driven than attempts to explain the significant proper motions of AGNs and quasars by more exotic effects.

\acknowledgements
This work is based on observations with the eROSITA
telescope onboard the SRG observatory. The SRG observatory was built by Roskosmos in the interests of the Russian Academy of Sciences represented by its Space Research Institute (IKI) within the framework of the Russian Federal Space Program, with the participation of the Deutsches Zentrum fuer Luft- und Raumfahrt (DLR). The
SRG/eROSITA X-ray telescope was built by a consortium of German Institutes led by MPE, and supported by DLR. The SRG spacecraft was designed, built, launched,
and is operated by the Lavochkin Association and its subcontractors. The science data are downlinked via the
Deep Space Network Antennae in Bear Lakes, Ussurijsk, and Baykonur, funded by Roskosmos. The eROSITA data used in this work were processed using the eSASS
software system developed by the German eROSITA consortium and the proprietary data reduction and analysis software developed by the Russian eROSITA Consortium. This research has made use of the SIMBAD database, operated at CDS, Strasbourg, France.
The authors are grateful to 
T\"UB\.{I}TAK, IKI, KFU, and the Academy of Sciences of the Tatarstan Republic  for partial support in use of RTT-150 (Russian-Turkish 1.5-m telescope in Antalya).

The study was supported by a grant from the Russian Science Foundation N 23-12-00292.


\bibliographystyle{astl}
\bibliography{cat.bib}

\begin{thebibliography}{}

\bibitem[\protect\citeauthoryear{{Biretta}
  {et~al.}}{1999}]{1999ApJ...520..621B}
J.~A. {Biretta}, W.~B. {Sparks}, and F. {Macchetto},
\newblock \apj, {\bf 520}, 621 (1999)

\bibitem[\protect\citeauthoryear{{Damineli}}{1996}]{1996ApJ...460L..49D}
A. {Damineli},
\newblock \apjl, {\bf 460}, L49 (1996)

\bibitem[\protect\citeauthoryear{{Gaia Collaboration}
  {et~al.}}{2021}]{GAIAEDR3}
{Gaia Collaboration}, A.~G.~A. {Brown}, A. {Vallenari}, T. {Prusti}, J.~H.~J.
  {de~Bruijne}, C. {Babusiaux}, et~al.,
\newblock \aap, {\bf 649}, A1 (2021)

\bibitem[\protect\citeauthoryear{{Gaia Collaboration}
  {et~al.}}{2022}]{Klioner2022}
{Gaia Collaboration}, S.~A. {Klioner}, L. {Lindegren}, F. {Mignard}, J.
  {Hernandez}, M. {Ramos-Lerate}, et~al.,
\newblock \aap, {\bf 667}, id.A148,31 (2022)

\bibitem[\protect\citeauthoryear{{Holl} {et~al.}}{2023}]{2023AA...674A..25H}
B. {Holl}, C. {Fabricius}, J. {Portell}, L. {Lindegren}, P. {Panuzzo}, M.
  {Bernet}, et~al.,
\newblock \aap, {\bf 674}, A25 (2023)

\bibitem[\protect\citeauthoryear{{Khamitov}
  {et~al.}}{2022}]{2022AstL...48..724K}
I.~M. {Khamitov}, I.~F. {Bikmaev}, M.~R. {Gilfanov}, R.~A. {Sunyaev}, P.~S.
  {Medvedev}, M.~A. {Gorbachev}, et~al.,
\newblock Astronomy Letters, {\bf 48}, 724 (2022)

\bibitem[\protect\citeauthoryear{{Kostrzewa-Rutkowska}
  {et~al.}}{2018}]{2018MNRAS.481..307K}
Z. {Kostrzewa-Rutkowska}, P.~G. {Jonker}, S.~T. {Hodgkin}, {\L}. {Wyrzykowski},
  M. {Fraser}, D.~L. {Harrison}, et~al.,
\newblock \mnras, {\bf 481}, 307 (2018)

\bibitem[\protect\citeauthoryear{{Kovalev} {et~al.}}{2017}]{2017AA...598L...1K}
Y.~Y. {Kovalev}, L. {Petrov}, and A.~V. {Plavin},
\newblock \aap, {\bf 598}, L1 (2017)

\bibitem[\protect\citeauthoryear{{Makarov} \&
  {Goldin}}{2016}]{2016ApJS..224...19M}
V.~V. {Makarov} and A. {Goldin},
\newblock \apjs, {\bf 224}, 19 (2016)

\bibitem[\protect\citeauthoryear{{Makarov} \&
  {Secrest}}{2022}]{2022ApJ...933...28M}
V.~V. {Makarov} and N.~J. {Secrest},
\newblock \apj, {\bf 933}, 28 (2022)

\bibitem[\protect\citeauthoryear{{Medvedev}
  {et~al.}}{2022}]{2022AstL...48..735M}
P.~S. {Medvedev}, M.~R. {Gilfanov}, S.~Y. {Sazonov}, R.~A. {Sunyaev}, and G.~A.
  {Khorunzhev},
\newblock Astronomy Letters, {\bf 48}, 735 (2022)

\bibitem[\protect\citeauthoryear{{Pierce} {et~al.}}{2023}]{2023MNRAS.522.1736P}
J.~C.~S. {Pierce}, C. {Tadhunter}, C. {Ramos Almeida}, P. {Bessiere}, J.~V.
  {Heaton}, S.~L. {Ellison}, et~al.,
\newblock \mnras, {\bf 522}, 1736 (2023)

\bibitem[\protect\citeauthoryear{{Plavin} {et~al.}}{2019}]{2019ApJ...871..143P}
A.~V. {Plavin}, Y.~Y. {Kovalev}, and L.~Y. {Petrov},
\newblock \apj, {\bf 871}, 143 (2019)

\bibitem[\protect\citeauthoryear{{Popovi{\'c}}
  {et~al.}}{2012}]{2012AA...538A.107P}
L.~{\v{C}}. {Popovi{\'c}}, P. {Jovanovi{\'c}}, M. {Stalevski}, S. {Anton},
  A.~H. {Andrei}, J. {Kova{\v{c}}evi{\'c}}, et~al.,
\newblock \aap, {\bf 538}, A107 (2012)

\bibitem[\protect\citeauthoryear{{Predehl} {et~al.}}{2021}]{2021AA...647A...1P}
P. {Predehl}, R. {Andritschke}, V. {Arefiev}, V. {Babyshkin}, O. {Batanov}, W.
  {Becker}, et~al.,
\newblock \aap, {\bf 647}, A1, 16 (2021)

\bibitem[\protect\citeauthoryear{{Souchay} {et~al.}}{2022}]{LCAQ5}
J. {Souchay}, N. {Secrest}, S. {Lambert}, N. {Zacharias}, F. {Taris}, C.
  {Barache}, et~al.,
\newblock \aap, {\bf 660}, A16, 1 (2022)

\bibitem[\protect\citeauthoryear{{Sunyaev} {et~al.}}{2021}]{2021AA...656A.132S}
R. {Sunyaev}, V. {Arefiev}, V. {Babyshkin}, A. {Bogomolov}, K. {Borisov}, M.
  {Buntov}, et~al.,
\newblock \aap, {\bf 656}, A132, 29 (2021)

\bibitem[\protect\citeauthoryear{{van Velzen}
  {et~al.}}{2021}]{2021ApJ...908....4V}
S. {van Velzen}, S. {Gezari}, E. {Hammerstein}, N. {Roth}, S. {Frederick}, C.
  {Ward}, et~al.,
\newblock \apj, {\bf 908}, 4 (2021)

\bibitem[\protect\citeauthoryear{{Wielen}}{1996}]{1996AA...314..679W}
R. {Wielen},
\newblock \aap, {\bf 314}, 679 (1996)

\end{thebibliography}

\onecolumn
\newpage
\twocolumn
\section{Appendix 1} 

\subsection{
Determination of the brightness center in the case of a transient event.}

We have considered the variation 
of the total brightness center in time and "GAIA data"$~$ of two objects: a) quasi-stationary and flared for a short time (a year) with a given light curve; b) quasi-stationary and with apparent linear motion. The objects are considered as a point-like.

In both cases, the study can be reduced to a one-dimensional model. In the picture plane, we refer the coordinate center to the position of the active nucleus, i.e., $X_{AGN}=0$, and orient the X axis away from the AGN toward the transient event (TE). Obviously, the brightness center with the transient event at a distance $X_T$ from the AGN position is determined as:
\begin{equation} \label{eq1}
X_C = \frac{X_T F_T}{F_T+\langle F_{AGN}\rangle+\xi(t)}, 
\end{equation}
where $\langle F_{AGN} \rangle$ is the average flux from AGN, $F_T$ is the flux from TE recorded at a certain moment of time, $\xi(t)$ is the stochastic variability of the core with respect to the mean value.

Consequently, there are two limit values of $X_C$. These are:

\begin{equation} 
X_C =  
\left\{
  \begin{array}{cc}
 X_T,       & \quad \text{if~ }  F_T \gg F_{AGN}\\
  0,         & \quad \text{if~ }  F_T \ll F_{AGN}
  \end{array}
  \right.
\label{eq2}
\end{equation}

The second limit value is reached rather quickly, e.g., for a supernova event or a tidal disruption event. Consequently, for known values of proper motion ($\mu$) measured in the GAIA catalog and the AGN flux estimatation, the lower value of the largest transient flux in the GAIA observation interval can be estimated in TE cases:
\begin{equation} \label{eq3}
    F_T=\frac{\mu t_{gaia}}{X_T-\mu t_{gaia}}F_{AGN},
\end{equation}
where $t_{gaia}$ is the time base of the GAIA eDR3 catalog, on the basis of which the proper motion of the sources was determined, i.e., 2.8 years.
Since both sources are at the same distance, the fluxes can be replaced by the luminosities of the sources in the formula \ref{eq3}. By adopting a conservative constraint on the maximum possible value of $X_T$ by the resolution element of the GAIA detector, i.e., 60 $mas$, the lower value of the transient luminosity was calculated. The results are summarized in tables \ref{cat_Sy1}-\ref{cat_galaxy} in the 9th column ($L_T$).

In addition to the proper motion parameter, the GAIA catalog also provides the astrometric noise parameter $\epsilon$ (astrometric\_excess\_ noise). By introducing a new variable $R$ as the ratio of TE fluxes to the average AGN flux and taking the ratio $\xi(t)$ to the sum of TE and AGN fluxes on the time base of the GAIA catalog to be less than unity, and in fact in the time interval when TE can be registered by Gaia detectors, (\ref{eq1}) takes the form:
\begin{equation} \label{eq4}
X_C=\frac{R}{R+1} X_T-\frac{R}{(R+1)^2}\xi^\prime(t)X_T,    
\end{equation}
where $\xi^\prime(t)=\frac{\xi(t)}{\langle F_{AGN}\rangle}$.  The second term of the sum in (\ref{eq4}), which has chaotic behavior, gives an additional small contribution to the astrometric noise, depending on the amplitude of the intrinsic relative variability of the core on the scale of the flare registration.

\subsection{The case of a quasi-stationary object and an object flared at a given distance.}

A simple model of the effect of a flare with a fast rise and exponential decay (FRED) profile that took place at some distance from the core ($X_T$) was considered. 
The fast rise is provided by a Gaussian: 
 \small
\begin{equation}
R_i = 
 R_0\times 
 \left\{
 \begin{array}{ll}
    e^{-(t_i-t_{peak})^2/2\sigma^2}   &\quad  \text{if~} t_i \leq  t_{peak} \\
    e^{-(t_i-t_{peak})/\alpha}         &\quad  \text{if~} t_i > t_{peak} 
  \end{array}
  \right.
\label{eq5}
\end{equation}

\normalsize
where $t_{peak}$ is the moment of maximum flare flux, $R_0$ is the ratio of TE and AGN fluxes at the moment $t_{peak}$, $\sigma$ is the standard deviation of the Gaussian, $\alpha$ is the exponential decay parameter, and $t_i$ are the moments of observations in the GAIA catalog, which are used in the model with some arbitrariness.  According to the GAIA observatory sky scanning parameters, each source is observed 70 times on average during a 5-year survey of the entire sky, i.e., at least one measurement per 30 days. Thus, for modeling it was assumed that at least 39 photometric  and positional measurements of extragalactic sources were carried out during the $t_{gaia}$.

\begin{figure}
    \centering
    \includegraphics[width=\columnwidth]{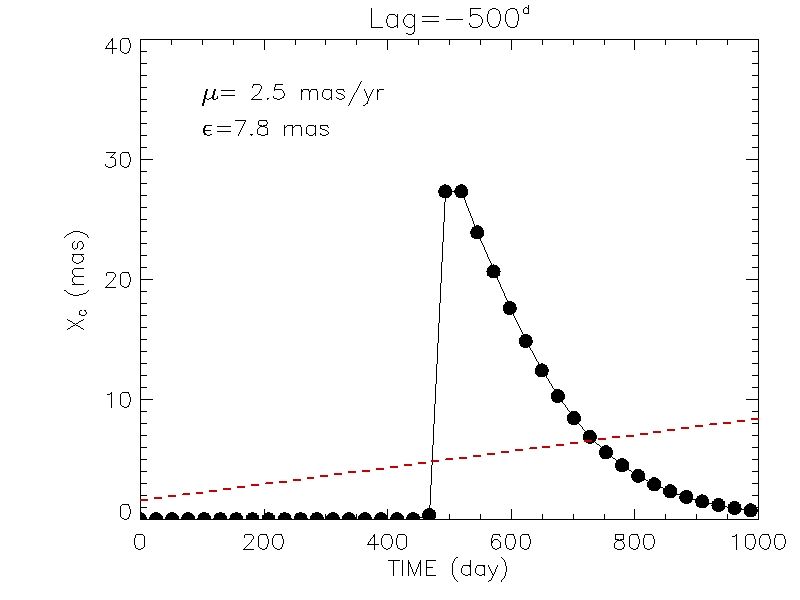}
    \caption{FRED-profile flare that has taken place 500 days after the beginning of the survey. The dots show simulated measurements of the photocenter of the AGN with the flare at a distance $X_T=60~mas$ from the nucleus ($R_0=1, \alpha=110^d, \sigma=10^d$). The zero value on the x-axis corresponds to the beginning of the GAIA survey. We present estimates of proper motion and astrometric noise obtained by linear approximation of all the measurements of the photocenter.}
    \label{fig:lagm500}
\end{figure}

Then, having estimated the position of the brightness center $X_C$ by the formulas (\ref{eq4}) and (\ref{eq5}), the proper motion $\mu$ and the standard deviation from the line $\epsilon$ are calculated for all points by linear interpolation ($\epsilon$ does not include the contribution from the stochastic variability of the core).  Depending on the time offset ($LAG$) of the flare maximum moment $t_{peak}$ compared to the beginning of the GAIA survey, these discrete measurements will correspond to different parts of the flare. Two cases can be distinguished depending on the sign of $LAG$. First, at negative values, the maximum of the flare is within the survey period (Fig.\ref{fig:lagm500}). Measurements of the brightness center before the flare will correspond to unbiased values regardless of the core variability, i.e., equal to zero. Second, with positive values, the maximum of the flare was before the beginning of the survey (Fig.\ref{fig:lagp100}) and all measurements of the brightness center would be displaced. These examples do not take into account the intrinsic variability of the core brightness, $X_T$ is 60 mas, $R_0=1$, and $\sigma=10^d$.

The considered fitting of the observed parameters ($\mu,\epsilon$) by FRED flare is a multiparametric problem. Indeed, decreasing the parameter $X_T$ or decreasing the parameter $R_0$ leads to a horizontal shift of the solutions towards a decrease in $\epsilon$ (Fig.\ref{fig:pm_epsi_FRED_XT},\ref{fig:pm_epsi_FRED_R0}). While decreasing the exponential decay parameter $\alpha$ leads to a vertical shift towards a decrease in the ratio of proper motions to astrometric noise (Fig.\ref{fig:pm_epsi_FRED_ALPHA}). The horizontal branches of the model estimates correspond to periods when the maximum of the transient event flux occurred before the beginning of the GAIA catalog, while the vertical part corresponds to moments when the flare took place within the time interval covered by the GAIA catalog. Smaller $\epsilon$ values on the horizontal branch correspond to larger $LAG$ parameters, and on the vertical curve smaller $2.8\mu/\epsilon$ values correspond to smaller $LAG$ values.

Besides the considered case with a quasi-stationary object and an object flaring at a given distance, there is a possible case of two quasi-stationary objects (distance between the objects $X_T$), in the neighborhood of one of which a transient event occurs. In this case, (\ref{eq1}) takes the form of:
\begin{equation}\label{eq11}
    X_C=\frac{X_TF_2+X_TF_T}{F_1+F_2+F_T}
\end{equation}
\begin{figure}
    \centering
    \includegraphics[width=\columnwidth]{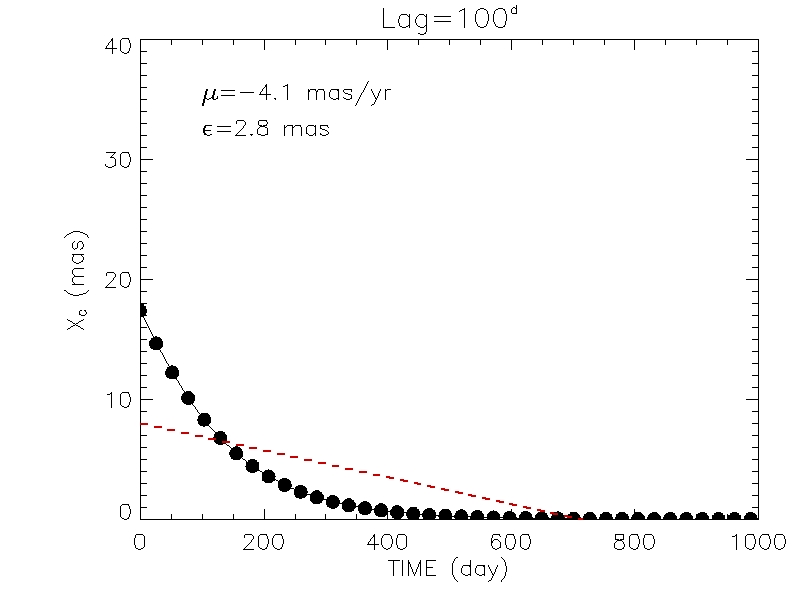}
    \caption{Photocenter variations in a case of FRED-profile flare that took place 100 days before the GAIA survey (model parameters: $X_T=60~mas, ~R_0=1,~\alpha=110^d,~ \sigma=10^d$). The negative value of the proper motion in the one-dimensional model corresponds to the direction from the flare toward the nucleus.}
    \label{fig:lagp100}
\end{figure}
where $F_1$ and $F_2$ are the fluxes of the first object and, respectively, the second object in the neighborhood of which the transient event took place. Introducing the variable $R=\frac{F_T}{F_1+F_2}$, (\ref{eq11}) takes the following form:
\begin{equation}
    X_C=\frac{X_0+X_TR}{1+R}
\end{equation}
where $X_0=\frac{X_TF_2}{F_1+F_2}$ is the mean coordinate of the photocenter of the double system in the absence of a transient event. The parameters $X_C$ and $R$ are GAIA observables from which we can estimate the distance between the objects and reconstruct the light curve of the transient event. 

\begin{figure}

    \centering
    \includegraphics[width=\columnwidth]{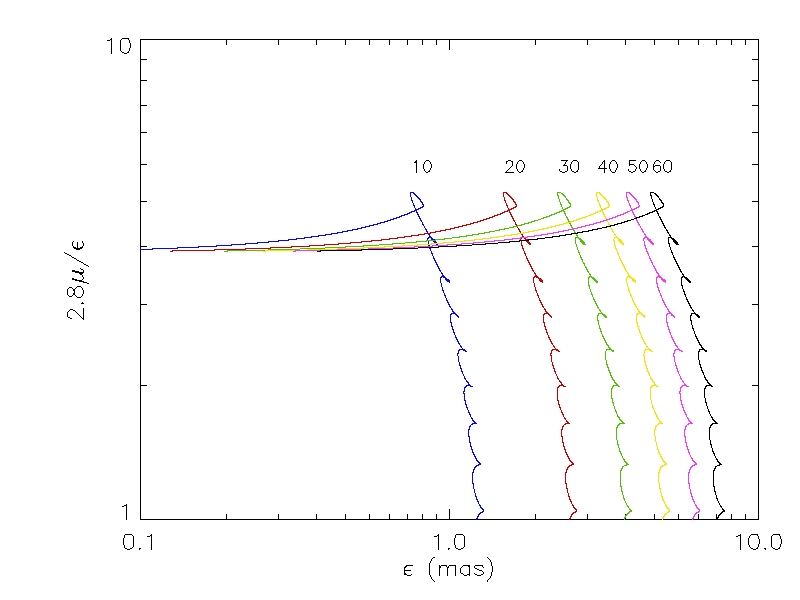}
    \caption{Variation of the model curves as a function of the distance of the transient from the nucleus $X_T$ ($R_0=1, \alpha=110^d, \sigma=10^d$). The numbers indicate the values of these distances in $mas$ units.}
    \label{fig:pm_epsi_FRED_XT}

\end{figure}

\begin{figure}
    \centering
    \includegraphics[width=\columnwidth]{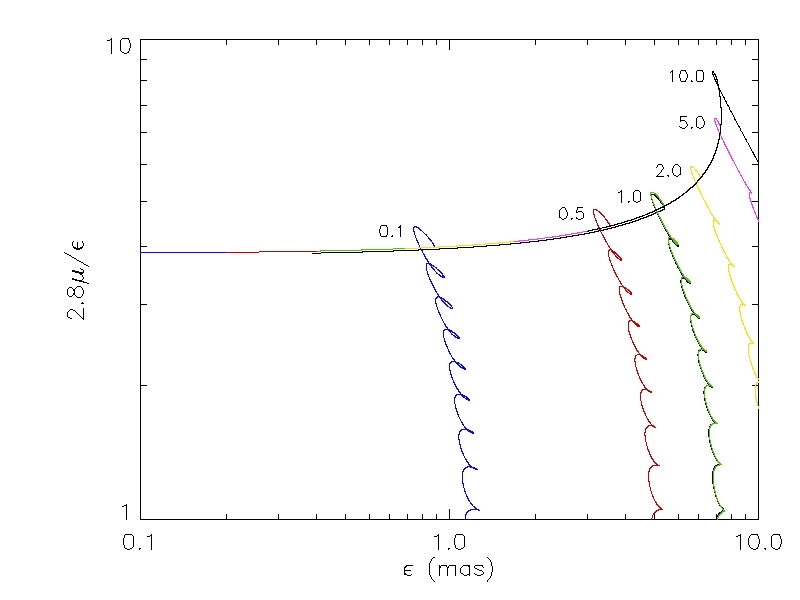}
    \caption{Variation of the model curves as a function of the ratio of the fluxes of the transient event to the flux from the AGN at time $t_{peak}$, $R_0$ ($X_T=60~mas, \alpha=110^d, \sigma=10^d$). The numbers indicate the values of $R_0$ corresponding to the curves.}
    \label{fig:pm_epsi_FRED_R0}
\end{figure}

\subsection{The case with a quasi-stationary object and a visible
linear motion of an variable brightness object.}

Similar cases take place with blazars, which demonstrate the presence of strong optical jets directed toward the observer at a small angle. In the calculation of the distribution ($\mu,\epsilon$), the coordinate $X_T$, in contrast to the previous case, varies linearly with time: $X_T=X_{T0}+\mu_{jet}t$, where $X_{T0}$ is the coordinate of TE at the initial moment of the GAIA catalog. Then, by the formula (\ref{eq4}) the apparent proper motion measured by Gaia is calculated as: $\mu=\frac{R}{(R+1)}\mu_{jet}+o(\xi^\prime)$.  And o-small is the linear part of the random parameter $\xi^\prime$, which includes the relative variability of the jet in addition to the relative variability of the core. The $\epsilon$ is a function of the ratio of TE to AGN fluxes ($R$), the relative internal variability of AGN and jet ($\xi^\prime$), the TE coordinate at the initial time of the GAIA catalog ($X_{T0}$), and the apparent proper motion of TE ($\mu_{jet}$). The calculation of this parameter is beyond the scope of this study.

\begin{figure}
    \centering
    \includegraphics[width=\columnwidth]{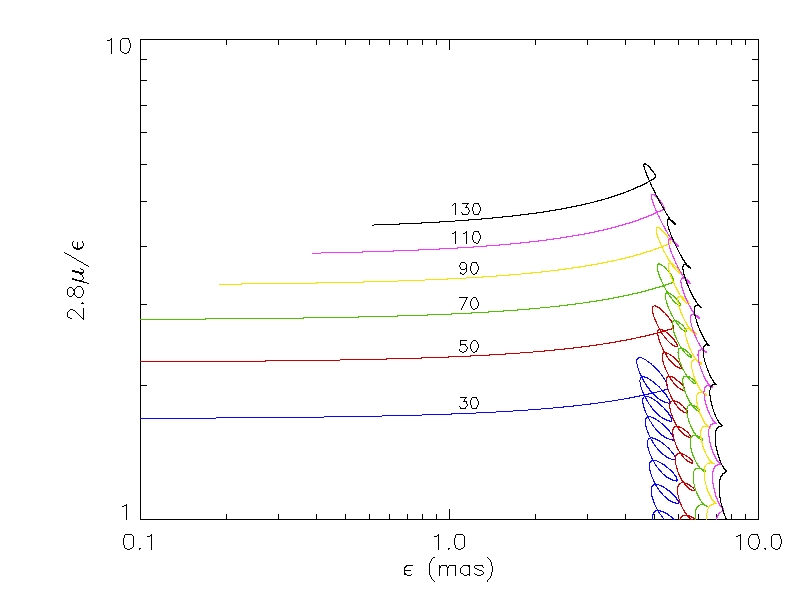}
    \caption{Variation of model curves depending on the exponential decay parameter $\alpha$ ($R_0=1,X_T=60~mas, \sigma=10^d$). The numbers indicate the values of these parameters in days.}
    \label{fig:pm_epsi_FRED_ALPHA}
\end{figure}

\scriptsize
\onecolumn 
\onehalfspacing
\begin{landscape}
\begin{longtable}{llcccclll}

 \caption{Catalog.\label{long}}\\

 \hline

 SRGe+          &	GAIA EDR3+  & RA          & DEC          &    $\mu$     &        z       &       Type   &     Simbad    &  Reference \\
               &                &             &               &   (mas/yr)   &                &               &              &             \\

 \hline
 \endfirsthead

 \multicolumn{9}{c}{Continuation of the table \ref{long}}\\
 \hline

 SRGe+          &	GAIA EDR3+  & RA          & DEC          &    $\mu$     &        z       &       Type   &     Simbad    &  Reference \\
               &                &             &               &   (mas/yr)   &                &               &              &             \\

\hline
\endhead
\endfoot

 \hline\hline

J000310.1+044455 & 2741905339356533504 & 00 03 10.02 & +04 44 56.2 & $ 1.94 \pm 0.51 $ & 0.058 & Sy2 & 2MASXJ00030999+0444564 & 2003A\&A...412...45P \\
J001124.6+380935 & 2877771750682930176 & 00 11 24.46 & +38 09 33.7 & $ 5.72 \pm 1.48 $ & 0.137 & AGN & None & 2022AstL...48..724K \\
J002848.8+145216 & 2780286129184910208 & 00 28 48.78 & +14 52 16.3 & $ 3.29 \pm 0.77 $ & 0.089 & Sy1 & 2MASXJ00284876+1452156 & 2012ApJS..203...21A \\
J002936.8-173833 & 2367248023501811840 & 00 29 36.79 & -17 38 30.4 & $ 3.85 \pm 0.30 $ & 0.054 & Galaxy & 2MASXJ00293678-1738306 & 2009MNRAS.399..683J \\
J003246.0-193931 & 2363442270160564480 & 00 32 46.03 & -19 39 30.4 & $ 3.12 \pm 0.85 $ & 0.054 & Galaxy & 2MASXJ00324605-1939303 & 2009MNRAS.399..683J \\
J003415.5-274813 & 2319321445518636928 & 00 34 15.47 & -27 48 12.7 & $ 3.69 \pm 0.44 $ & 0.005 & Galaxy SF & NGC150 & 2012ApJS..203...21A \\
J003826.9-000039 & 2543175415864489472 & 00 38 26.67 & -00 00 42.7 & $ 2.99 \pm 0.68 $ & 0.045 & Galaxy & MCG+00-02-101 & 2012ApJS..203...21A \\
J004447.5+152910 & 2781106124341566976 & 00 44 47.34 & +15 29 11.9 & $ 6.51 \pm 1.17 $ & 0.227 & Sy1 & SDSSJ004447.33+152911.8 & 2012ApJS..203...21A \\
\hline\hline
\end{longtable}

\end{landscape}
\singlespacing
\normalsize
\onecolumn
\begin{longtable}{llccccccc}

 \caption{Catalog of extragalactic sources with significant proper motions - Seyfert galaxies of type 1.\label{cat_Sy1}}\\

 \hline

   RA (J2000) & DEC (J2000) &  G    &    $\mu$   &  $\tau$  & $X_v$ &  $L_X \times10^{42}$ & z & $L_T \times10^{42}$\\
            &             & (mag) &  (mas/yr)  &    &        &  ($erg/s$)          &   & ($erg/s$)                    \\

 \hline
 \endfirsthead

 \multicolumn{9}{c}{Continuation of table \ref{cat_Sy1} - Seyfert galaxies of type 1.}\\
 \hline

  RA (J2000) & DEC (J2000) &  G    &    $\mu$   &  $\tau$  & $X_v$ &  $L_X \times10^{42}$ & z & $L_T \times10^{42}$\\
            &             & (mag) &  (mas/yr)  &    &        &  ($erg/s$)          &   & ($erg/s$)                    \\

\hline
\endhead
\endfoot

 \hline\hline
           
11 05 04.21 & +50 59 49.9 & 19.86 & 9.43 & -0.3 & 4.9 & 3.3 & 0.118 & 5.2    \\
14 40 25.84 & +33 27 02.6 & 20.82 & 9.41 & -0.7 & 2.7 & 3.4 & 0.275 & 13.5    \\
14 54 25.48 & +46 45 24.1 & 18.99 & 8.99 & 0.1 & 5.2 & 5.7 & 0.069 & 3.3    \\
13 56 20.70 & +26 43 54.4 & 19.66 & 8.86 & 0.3 & 24.2 & 3.7 & 0.062 & 1.3    \\
11 02 41.47 & +42 06 51.9 & 19.35 & 7.02 & -0.2 & 3.1 & 2.8 & 0.075 & 2.1    \\
13 31 06.88 & +51 09 30.9 & 20.15 & 6.85 & -0.1 & 1.8 & 9.3 & 0.179 & 5.5    \\
00 44 47.34 & +15 29 11.9 & 20.22 & 6.51 & -0.4 & 5.2 & 7.6 & 0.227 & 8.4    \\
14 39 36.35 & +44 55 04.3 & 19.29 & 6.03 & 0.1 & 5.2 & 8.4 & 0.088 & 2.6    \\
\hline\hline
\end{longtable}

\begin{longtable}{llccccccc}

 \caption{Catalog of extragalactic sources with significant proper motions - Seyfert galaxies of type 2.\label{cat_Sy2}}\\

 \hline

    RA (J2000) & DEC (J2000) &  G    &    $\mu$   &  $\tau$  & $X_v$ &  $L_X \times10^{42}$ & z & $L_T \times10^{42}$\\
            &             & (mag) &  (mas/yr)  &    &        &  ($erg/s$)          &   & ($erg/s$)                    \\

 \hline
 \endfirsthead

 \multicolumn{9}{c}{Continuation of table \ref{cat_Sy2} - Seyfert galaxies of type 2.}\\
 \hline

    RA (J2000) & DEC (J2000) &  G    &    $\mu$   &  $\tau$  & $X_v$ &  $L_X \times10^{42}$ & z & $L_T \times10^{42}$\\
            &             & (mag) &  (mas/yr)  &    &        &  ($erg/s$)          &   & ($erg/s$)                    \\

\hline
\endhead
\endfoot

 \hline\hline
           
16 08 51.07 & +29 57 15.0 & 19.91 & 9.98 & -0.9 & 0.0 & 0.1 & 0.048 & 0.8    \\
16 53 15.06 & +23 49 42.9 & 19.58 & 8.61 & -1.1 & 0.0 & 0.6 & 0.103 & 4.7    \\
14 16 07.75 & +35 20 37.8 & 17.99 & 6.75 & -1.4 & 4.7 & 0.02 & 0.013 & 0.2    \\
12 46 17.34 & +28 20 33.9 & 19.13 & 5.35 & -0.6 & 5.9 & 2.4 & 0.100 & 3.1    \\
13 15 17.26 & +44 24 25.5 & 18.71 & 5.33 & -0.8 & 6.1 & 0.2 & 0.035 & 0.5    \\
06 52 36.99 & +45 46 50.2 & 18.08 & 4.90 & -1.5 & 2.6 & 0.03 & 0.021 & 0.3    \\
14 23 07.51 & +28 35 42.3 & 18.21 & 4.25 & -0.9 & 3.7 & 0.2 & 0.029 & 0.4    \\
14 08 45.73 & +35 32 18.6 & 19.27 & 3.60 & -1.3 & 0.0 & 1.3 & 0.166 & 5.1    \\
\hline\hline
\end{longtable}

\begin{longtable}{llccccccc}

 \caption{Catalog of extragalactic sources with significant proper motions - AGN of undetermined type.\label{cat_AGN}}\\

 \hline

 RA (J2000) & DEC (J2000) &  G    &    $\mu$   &  $\tau$  & $X_v$ &  $L_X \times10^{42}$ & z & $L_T \times10^{42}$\\
            &             & (mag) &  (mas/yr)  &    &        &  ($erg/s$)          &   & ($erg/s$)                    \\

 \hline
 \endfirsthead

 \multicolumn{9}{c}{Continuation of the table \ref{cat_AGN} - AGN of undetermined type.}\\
 \hline

RA (J2000) & DEC (J2000) &  G    &    $\mu$   &  $\tau$  & $X_v$ &  $L_X \times10^{42}$ & z & $L_T \times10^{42}$\\
            &             & (mag) &  (mas/yr)  &    &        &  ($erg/s$)          &   & ($erg/s$)                    \\

\hline
\endhead
\endfoot

 \hline\hline

14 19 43.23 & +49 14 11.9 & 17.35 & 6.10 & -1.8 & 2.4 & 0.04 & 0.026 & 1.0  \\
14 43 31.26 & +49 23 35.2 & 18.93 & 5.82 & -0.2 & 5.7 & 0.6 & 0.030 & 0.4  \\
00 11 24.46 & +38 09 33.7 & 19.92 & 5.72 & -0.4 & 4.4 & 3.8 & 0.137 & 3.5  \\
06 23 29.08 & +69 12 32.6 & 19.93 & 4.54 & -0.6 & 2.9 & 0.3 & 0.055 & 0.4  \\
05 38 23.46 & +79 35 12.7 & 17.53 & 4.27 & -2.1 & 0.0 & 0.007 & 0.015 & 0.2  \\
15 39 37.06 & +59 19 55.2 & 19.26 & 4.13 & -0.8 & 1.6 & 0.009 & 0.008 & 0.01  \\
01 20 48.01 & -08 29 18.4 & 18.69 & 4.10 & -0.5 & 2.6 & 0.5 & 0.034 & 0.3   \\
19 00 50.52 & +28 46 16.4 & 18.95 & 4.10 & -0.4 & 9.3 & 0.08 & 0.014 & 0.05  \\
\hline\hline
\end{longtable}

\begin{longtable}{llccccccc}

 \caption{Catalog of extragalactic sources with significant proper motions - LINER-type AGNs.\label{cat_LINER}}\\

 \hline

 RA (J2000) & DEC (J2000) &  G    &    $\mu$   &  $\tau$  & $X_v$ &  $L_X \times10^{42}$ & z & $L_T \times10^{42}$\\
            &             & (mag) &  (mas/yr)  &    &        &  ($erg/s$)          &   & ($erg/s$)                    \\

 \hline
 \endfirsthead

 \multicolumn{9}{c}{Continuation of the table \ref{cat_LINER}}\\
 \hline

 RA (J2000) & DEC (J2000) &  G    &    $\mu$   &  $\tau$  & $X_v$ &  $L_X \times10^{42}$ & z & $L_T \times10^{42}$\\
            &             & (mag) &  (mas/yr)  &    &        &  ($erg/s$)          &   & ($erg/s$)                    \\
 
\hline
\endhead
\endfoot

 \hline\hline
          
09 11 18.96 & +46 23 07.2 & 19.43 & 7.93 & -0.3 & 8.1 & 3.8 & 0.103 & 4.4   \\
16 13 18.50 & +36 16 21.9 & 19.80 & 7.71 & -1.2 & 2.4 & 0.2 & 0.082 & 1.7   \\
09 05 54.50 & +47 10 45.5 & 18.74 & 4.08 & -1.0 & 3.2 & 0.08 & 0.027 & 0.2   \\
15 53 40.36 & +43 44 04.9 & 19.25 & 4.06 & -0.2 & 2.4 & 0.8 & 0.040 & 0.3   \\
12 02 15.46 & +44 03 19.7 & 19.65 & 3.80 & -0.1 & 8.9 & 4.3 & 0.101 & 1.2   \\
16 51 29.52 & +19 24 59.9 & 18.30 & 3.61 & -0.8 & 3.7 & 0.1 & 0.023 & 0.2   \\
12 10 49.61 & +39 28 22.1 & 18.32 & 3.54 & -0.4 & 24.8 & 0.4 & 0.022 & 0.2   \\
14 33 18.48 & +34 44 04.4 & 18.93 & 3.06 & -0.5 & 1.9 & 0.3 & 0.034 & 0.2   \\
13 43 30.01 & +51 02 04.1 & 19.70 & 2.93 & 0.3 & 4.1 & 17.7 & 0.131 & 1.4   \\
11 22 17.90 & +59 04 28.3 & 17.74 & 1.88 & -0.9 & 10.2 & 0.008 & 0.005 & 0.006   \\
17 03 15.72 & +31 27 28.6 & 17.04 & 1.67 & -1.6 & 1.7 & 0.2 & 0.034 & 0.6   \\

 \hline\hline
\end{longtable}

\begin{longtable}{llccccccc}

 \caption{Catalog of extragalactic sources with significant proper motions - quasars.\label{cat_qso}}\\

 \hline

   RA (J2000) & DEC (J2000) &  G    &    $\mu$   &  $\tau$  & $X_v$ &  $L_X \times10^{42}$ & z & $L_T \times10^{42}$\\
            &             & (mag) &  (mas/yr)  &    &        &  ($erg/s$)          &   & ($erg/s$)                    \\

 \hline
 \endfirsthead

 \multicolumn{9}{c}{Continuation of the table \ref{cat_qso} - quasars.}\\
 \hline

 RA (J2000) & DEC (J2000) &  G    &    $\mu$   &  $\tau$  & $X_v$ &  $L_X \times10^{42}$ & z & $L_T \times10^{42}$\\
            &             & (mag) &  (mas/yr)  &    &        &  ($erg/s$)          &   & ($erg/s$)                    \\

\hline
\endhead
\endfoot

 \hline\hline
           
13 11 18.54 & +46 35 02.3 & 19.45 & 14.94 & -0.3 & 2.4 & 28.4 & 0.271 & 130.3    \\
01 44 17.27 & +31 40 03.3 & 20.09 & 9.32 & -0.2 & 6.6 & 3.5 & 0.124 & 4.2    \\
23 53 30.74 & +11 52 52.5 & 20.06 & 4.64 & 0.1 & 2.2 & 29.8 & 0.235 & 6.6    \\
00 50 55.69 & +29 33 28.1 & 19.63 & 4.56 & -0.5 & 5.8 & 3.3 & 0.136 & 2.8    \\
01 34 35.91 & +02 28 39.9 & 19.32 & 3.51 & -0.2 & 8.6 & 15.5 & 0.177 & 4.8    \\
14 49 24.44 & +32 18 16.2 & 18.37 & 3.29 & -0.3 & 5.9 & 2.6 & 0.058 & 1.0    \\
15 09 08.76 & +09 02 21.0 & 18.31 & 2.65 & -1.5 & 0.0 & 0.1 & 0.044 & 0.6    \\
14 37 01.50 & +26 40 19.2 & 19.06 & 2.08 & 0.0 & 3.1 & 56.3 & 0.218 & 6.0    \\
21 49 03.97 & -01 41 11.8 & 18.24 & 1.44 & 0.0 & 3.3 & 5.7 & 0.053 & 0.4    \\
17 55 05.61 & +65 19 55.0 & 17.32 & 0.73 & -0.1 & 2.1 & 21.2 & 0.079 & 0.9    \\
22 27 19.04 & +40 05 50.3 & 17.42 & 0.64 & -1.0 & 6.4 & 2.0 & 0.068 & 0.6    \\

 \hline\hline
\end{longtable}
\begin{longtable}{llccccccc}

 \caption{Catalog of extragalactic sources with significant proper motions - blazars.\label{cat_BLL}}\\

 \hline


RA (J2000) & DEC (J2000) &  G    &    $\mu$   &  $\tau$  & $X_v$ &  $L_X \times10^{42}$ & z & $L_T \times10^{42}$\\
            &             & (mag) &  (mas/yr)  &    &        &  ($erg/s$)          &   & ($erg/s$)                    \\

 \hline
 \endfirsthead

 \multicolumn{9}{c}{Continuation of table \ref{cat_BLL} - blazars.}\\
 \hline


RA (J2000) & DEC (J2000) &  G    &    $\mu$   &  $\tau$  & $X_v$ &  $L_X \times10^{42}$ & z & $L_T \times10^{42}$\\
            &             & (mag) &  (mas/yr)  &    &        &  ($erg/s$)          &   & ($erg/s$)                    \\

\hline
\endhead
\endfoot

 \hline\hline
          
16 40 58.89 & +11 44 04.2 & 18.00 & 2.05 & -0.4 & 1.6 & 6.0 & 0.078 & 1.6 \\
07 52 44.21 & +45 56 57.4 & 18.17 & 1.29 & -0.1 & 2.1 & 4.6 & 0.052 & 0.4 \\

 \hline\hline
\end{longtable}

\begin{longtable}{llccccccc}

 \caption{Catalog of extragalactic sources with significant proper motions - radio galaxies.\label{cat_radio}}\\

 \hline

  RA (J2000) & DEC (J2000) &  G    &    $\mu$   &  $\tau$  & $X_v$ &  $L_X \times10^{42}$ & z & $L_T \times10^{42}$\\
            &             & (mag) &  (mas/yr)  &    &        &  ($erg/s$)          &   & ($erg/s$)                    \\

 \hline
 \endfirsthead

 \multicolumn{9}{c}{Continuation of the table \ref{cat_radio} - radio galaxies.}\\
 \hline

  RA (J2000) & DEC (J2000) &  G    &    $\mu$   &  $\tau$  & $X_v$ &  $L_X \times10^{42}$ & z & $L_T \times10^{42}$\\
            &             & (mag) &  (mas/yr)  &    &        &  ($erg/s$)          &   & ($erg/s$)                    \\

\hline
\endhead
\endfoot

 \hline\hline
           
12 43 11.22 & +73 15 59.3 & 19.06 & 4.96 & -0.2 & 2.4 & 3.3 & 0.075 & 1.6    \\
01 19 59.60 & +14 47 10.4 & 17.11 & 4.83 & -1.9 & 3.1 & 0.01 & 0.014 & 0.3    \\
15 22 40.24 & +31 08 56.7 & 19.02 & 4.32 & -0.9 & 5.2 & 1.9 & 0.116 & 3.8    \\
23 36 04.96 & -31 34 51.0 & 19.06 & 3.45 & -0.4 & 1.5 & 1.3 & 0.062 & 0.6    \\

 \hline\hline
\end{longtable}
\begin{longtable}{llccccccc}

 \caption{Catalog of extragalactic sources with significant proper motions - star-forming galaxies.\label{cat_SFgalaxy}}\\

 \hline

  RA (J2000) & DEC (J2000) &  G    &    $\mu$   &  $\tau$  & $X_v$ &  $L_X \times10^{42}$ & z & $L_T \times10^{42}$\\
            &             & (mag) &  (mas/yr)  &    &        &  ($erg/s$)          &   & ($erg/s$)                    \\

 \hline
 \endfirsthead

 \multicolumn{9}{c}{Continuation of the table \ref{cat_SFgalaxy} - star-forming galaxies.}\\
 \hline

   RA (J2000) & DEC (J2000) &  G    &    $\mu$   &  $\tau$  & $X_v$ &  $L_X \times10^{42}$ & z & $L_T \times10^{42}$\\
            &             & (mag) &  (mas/yr)  &    &        &  ($erg/s$)          &   & ($erg/s$)                    \\

\hline
\endhead
\endfoot

 \hline\hline
15 17 22.00 & +46 58 12.9 & 20.61 & 17.63 & -0.5 & 3.1 & 2.2 & 0.154 & 27.8  \\      
15 13 45.76 & +31 11 25.1 & 20.48 & 7.37 & -0.4 & 5.1 & 0.5 & 0.072 & 0.7    \\
21 24 04.81 & -16 41 48.1 & 19.17 & 7.01 & -0.8 & 0.0 & 0.2 & 0.036 & 0.5    \\
12 35 02.66 & +66 22 33.5 & 18.22 & 5.41 & -1.7 & 0.0 & 0.09 & 0.047 & 1.5    \\
14 54 04.58 & +36 33 20.5 & 19.79 & 5.20 & -0.8 & 5.6 & 0.4 & 0.071 & 0.7    \\
13 36 50.95 & +36 50 18.0 & 18.76 & 5.15 & -0.6 & 1.6 & 1.2 & 0.062 & 1.5    \\
15 14 44.82 & +26 54 32.5 & 19.96 & 5.12 & -0.3 & 5.0 & 4.0 & 0.138 & 2.5    \\
14 42 27.61 & +55 58 46.4 & 18.61 & 4.61 & -1.2 & 1.6 & 0.5 & 0.077 & 2.2    \\
\hline\hline
\end{longtable}
\begin{longtable}{llccccccc}

 \caption{Catalog of extragalactic sources with significant proper motions - galaxies with undetermined type.\label{cat_galaxy}}\\

 \hline

 RA (J2000) & DEC (J2000) &  G    &    $\mu$   &  $\tau$  & $X_v$ &  $L_X \times10^{42}$ & z & $L_T \times10^{42}$\\
            &             & (mag) &  (mas/yr)  &    &        &  ($erg/s$)          &   & ($erg/s$)                    \\

 \hline
 \endfirsthead

 \multicolumn{9}{c}{Continuation of the table \ref{cat_galaxy} - galaxies with undetermined type.}\\
 \hline

RA (J2000) & DEC (J2000) &  G    &    $\mu$   &  $\tau$  & $X_v$ &  $L_X \times10^{42}$ & z & $L_T \times10^{42}$\\
            &             & (mag) &  (mas/yr)  &    &        &  ($erg/s$)          &   & ($erg/s$)                    \\

\hline
\endhead
\endfoot

 \hline\hline

04 31 22.51 & +60 55 00.2 & 18.49 & 7.50 & -1.5 & 0.0 & 0.010 & 0.015 & 0.2    \\
05 08 09.16 & +70 05 30.1 & 19.40 & 5.93 & -0.3 & 2.6 & 0.2 & 0.028 & 0.2   \\
16 24 33.47 & +55 54 52.6 & 18.54 & 5.73 & -1.7 & 2.1 & 0.03 & 0.031 & 0.6   \\
16 52 51.42 & +17 26 51.0 & 19.16 & 5.68 & 0.5 & 4.7 & 2.7 & 0.034 & 0.3   \\
16 32 20.70 & +64 10 53.5 & 19.15 & 5.31 & -1.4 & 2.5 & 0.1 & 0.065 & 1.2   \\
23 35 16.07 & -08 57 23.5 & 20.04 & 5.23 & -0.7 & 0.0 & 0.6 & 0.086 & 1.0   \\
19 28 52.31 & -25 16 38.7 & 19.17 & 5.20 & -0.9 & 2.0 & 1.4 & 0.108 & 3.5   \\
05 13 16.43 & +66 27 50.2 & 19.06 & 5.12 & -0.2 & 2.4 & 0.1 & 0.015 & 0.05   \\
\hline\hline
\end{longtable}

\twocolumn

\end{document}